\documentclass[12pt,epsf]{article}
\pdfoutput=1
\usepackage{graphicx}
\usepackage{amsmath}
\usepackage{amsfonts}
\usepackage{amssymb}
\usepackage{cite}
\begin{document}

\def\a{\alpha}
\def\b{\beta}
\def\c{\varepsilon}
\def\d{\delta}
\def\e{\epsilon}
\def\f{\phi}
\def\g{\gamma}
\def\h{\theta}
\def\k{\kappa}
\def\l{\lambda}
\def\m{\mu}
\def\n{\nu}
\def\p{\psi}
\def\q{\partial}
\def\r{\rho}
\def\s{\sigma}
\def\t{\tau}
\def\u{\upsilon}
\def\v{\varphi}
\def\w{\omega}
\def\x{\xi}
\def\y{\eta}
\def\z{\zeta}
\def\D{\Delta}
\def\G{\Gamma}
\def\H{\Theta}
\def\L{\Lambda}
\def\F{\Phi}
\def\P{\Psi}
\def\S{\Sigma}

\def\o{\over}
\def\beq{\begin{eqnarray}}
\def\eeq{\end{eqnarray}}
\newcommand{\gsim}{ \mathop{}_{\textstyle \sim}^{\textstyle >} }
\newcommand{\lsim}{ \mathop{}_{\textstyle \sim}^{\textstyle <} }
\newcommand{\vev}[1]{ \left\langle {#1} \right\rangle }
\newcommand{\bra}[1]{ \langle {#1} | }
\newcommand{\ket}[1]{ | {#1} \rangle }
\newcommand{\EV}{ {\rm eV} }
\newcommand{\KEV}{ {\rm keV} }
\newcommand{\MEV}{ {\rm MeV} }
\newcommand{\GEV}{ {\rm GeV} }
\newcommand{\TEV}{ {\rm TeV} }
\newcommand{\be}{\begin{equation}}  
\newcommand{\ee}{\end{equation}}  
\newcommand{\ol}[1]{\overline{#1}}
\newcommand{\hc}{+\,\mathrm{h.c.}}
\newcommand{\into}{\ensuremath{\,\rightarrow\,}}
\newcommand{\U}{\operatorname{U}}
\def\diag{\mathop{\rm diag}\nolimits}
\def\Spin{\mathop{\rm Spin}}
\def\SO{\mathop{\rm SO}}
\def\O{\mathop{\rm O}}
\def\SU{\mathop{\rm SU}}
\def\Sp{\mathop{\rm Sp}}
\def\SL{\mathop{\rm SL}}
\def\tr{\mathop{\rm tr}}
\def\mpl{M_{PL}}

\def\IJMP{Int.~J.~Mod.~Phys. }
\def\MPL{Mod.~Phys.~Lett. }
\def\NP{Nucl.~Phys. }
\def\PL{Phys.~Lett. }
\def\PR{Phys.~Rev. }
\def\PRL{Phys.~Rev.~Lett. }
\def\PTP{Prog.~Theor.~Phys. }
\def\ZP{Z.~Phys. }


\baselineskip 0.7cm

\begin{titlepage}

\begin{flushright}
DESY 13-042\\
IPMU 13-0055
\end{flushright}

\vskip 1.35cm
\begin{center}
{\bf\Large Focus point gauge mediation\\ in product group unification

}

\vskip 1.2cm
Felix Br\"ummer$^1$, Masahiro Ibe$^{2,3}$ and Tsutomu T. Yanagida$^2$
\vskip 0.4cm
$^1${\it Deutsches Elektronen-Synchrotron DESY, D-22603 Hamburg, Germany}\\
$^2${\it Kavli IPMU, TODIAS, University of Tokyo, Kashiwa 277-8583, Japan}\\
$^3${\it ICRR, University of Tokyo, Kashiwa 277-8582, Japan}
\vskip 1.5cm

\abstract{\noindent
In certain models of gauge-mediated supersymmetry breaking with messenger fields in
incomplete GUT multiplets, the radiative corrections to the Higgs potential
cancel out during renormalization group running. This allows for relatively heavy
superpartners and for a 125 GeV Higgs while the fine-tuning remains modest.
In this paper, we show that such gauge mediation models with ``focus point'' 
behaviour can be naturally embedded into a model of $\SU(5)\times \U(3)$ product 
group unification.
}
\end{center}
\end{titlepage}

\setcounter{page}{2}

\section{Introduction}

If low-energy supersymmetry is realized in nature, the LHC results of the last
two years point towards a rather high superpartner mass scale, perhaps in the
range of several TeV.
However, obtaining an electroweak symmetry breaking 
scale which is an order of magnitude or more below the superpartner mass scale 
requires significant fine-tuning. This is the well-known little hierarchy
problem of supersymmetry.

Models with non-unified gaugino masses have recently been argued to alleviate
the supersymmetric fine-tuning problem in the MSSM
\cite{Abe:2007kf,Horton:2009ed,Brummer:2011yd,Brummer:2012zc,Younkin:2012ui,
Antusch:2012gv,Yanagida:2013ah}. For suitable ``focus point'' ratios of gaugino
masses, the radiative corrections to the Higgs potential cancel out during
renormalization group running, yielding an electroweak scale which is much
smaller than the typical scale of soft SUSY-breaking masses.\footnote{This 
is similar in spirit to the original focus point SUSY scenario
\cite{Chan:1997bi,Feng:1999mn,Feng:1999zg},
where instead the scalar soft mass contributions to the electroweak scale cancel
during the renormalization group evolution.}

High-scale gauge mediation with messenger fields in incomplete GUT multiplets
can naturally realize suitable non-universal soft mass ratios 
\cite{Brummer:2012zc}. In terms of the messenger indices $N_2$ 
and $N_3$ for pairs of fundamental $\SU(2)_{\rm L}$ and $\SU(3)_{\rm C}$
messenger multiplets, the favourable models tend to have a ratio around 
$N_2\,:\,N_3\;\approx\; 5\,:\,2$ (if the mediation scale is close to the GUT
scale, and if $\tan\beta$ is large). Models with such an exotic field content 
may be obtained from higher-dimensional orbifold GUTs, or from related heterotic 
string constructions \cite{Brummer:2011yd}. In the present letter we are proposing 
an example within the more conventional setting of four-dimensional field theory.

Of course any successful GUT model needs to somehow accommodate incomplete GUT 
multiplets, in order to solve the doublet-triplet splitting problem in the
Higgs sector. Models of product group unification (PGU) 
\cite{Yanagida:1994vq,Hisano:1995hc,Izawa:1997he} achieve this  by extending
the unified gauge group to, for instance,
$\SU(5)\times\SU(3)_{\rm H}\times\U(1)_{\rm H}$. Below the GUT scale, colour
$\SU(3)_{\rm C}$ is obtained as the diagonal subgroup of $\SU(3)_{\rm H}$ and
the Georgi-Glashow embedded $\SU(3)\subset\SU(5)$, and hypercharge is similarly 
a linear combination of $\U(1)_{\rm H}$ and $\U(1)\subset\SU(5)$. In such 
models the doublet-triplet splitting problem is easily solved, and also 
additional incomplete GUT multiplets can be accommodated straightforwardly.

In PGU models the gauge couplings need not unify. In fact, the theory will 
become non-perturbative immediately above the GUT scale unless the gauge 
couplings are prevented from unifying. In order to allow for a cutoff 
scale which is not at $M_{\rm GUT}$ but significantly higher, one needs to
ensure that $g_2$ is the largest among the Standard Model gauge couplings at 
the GUT scale. This is most easily achieved by adding some additional pairs 
of $\SU(2)_{\rm L}$ doublets with intermediate-scale masses.

Intriguingly, PGU and focus point gauge mediation are seen to complement each
other. On the one hand, for a focus point-like cancellation in gauge mediation, 
the model needs to contain significantly more weak doublet messengers than
colour triplet messengers. On the other hand, in order to maximize the cutoff
scale in PGU one needs to deflect the renormalization group running of $g_2$ 
relative to the other couplings by adding extra vector-like states, again with 
significantly more weak doublets than colour triplets. 

In the present letter we exploit this observation, by constructing a PGU model 
with a number of incomplete GUT multiplets with masses below the GUT scale. 
These will act as gauge mediation messengers, inducing non-unified soft term
ratios favourable for naturalness, while at the same time ensuring that the 
theory remains valid perturbatively at energies above $M_{\rm GUT}$.

Our model predicts the lightest Higgs mass to be compatible with the recent 
LHC discovery, thanks to large radiative corrections from multi-TeV soft terms, 
and evades the LHC limits on squark and gluino masses. The gauge couplings will 
become non-perturbative at a cutoff scale $M_*$, which is about an order of 
magnitude larger than $M_{\rm GUT}$ (but still below the Planck scale).

Given that the present LHC bounds on gluino and squark masses are around 1.5 TeV,
large parts of the remaining MSSM parameter space are fine-tuned to the level
of a permille or worse when taking all soft terms as independent. However,
one should bear in mind that the degree of fine-tuning actually depends on the 
high-scale model in which the soft terms are generated. It is therefore important
to search for models in which the fine-tuning is reduced. This can be achieved in
models which predict suitable relations between the soft terms, thus lowering the 
sensitivity of the resulting electroweak scale to the actual fundamental parameters.

In our model, characterized by the messenger indices $N_2=17$ and $N_3=7$, this
is precisely what happens. With this messenger field content, the predicted ratio
of gaugino masses is such that the electroweak scale is no longer very sensitive
to the actual soft mass scale. Therefore, despite the soft masses being of the 
order of a few TeV, the fine-tuning is modest compared to generic MSSM models with 
similarly heavy superpartners. The residual fine-tuning is of the order of a 
percent.

\section{Field content and evolution of couplings} 

We start by briefly reviewing the main properties of the $\SU(5)\times\U(3)$ PGU
model; for more details see e.g.~\cite{Yanagida:1994vq,Hisano:1995hc,
Izawa:1997he}. Consider a supersymmetric GUT with gauge group
$\SU(5)\times\SU(3)_{\rm
H}\times\U(1)_{\rm H}$. There are three generations of Standard Model matter
fields in the ${\bf 10}\oplus\ol{\bf 5}$ and a pair of Higgs fields $H,\ol H$ in
the ${\bf 5}\oplus\ol{\bf 5}$ of $\SU(5)$, all of which are uncharged under
$\SU(3)_{\rm H}\times\U(1)_{\rm H}$. Bi-fundamental fields transforming as
$Y=({\bf 5},\ol{\bf 3})$ and $\ol Y=(\ol{\bf 5},{\bf 3})$ under
$\SU(5)\times\SU(3)_{\rm H}$ acquire vacuum expectation values at the GUT
breaking
scale $M_{\rm GUT}$, thus breaking $\SU(5)\times\SU(3)_{\rm H}\times\U(1)_{\rm
H}\into\SU(3)_{\rm C}\times\SU(2)_{\rm L}\times\U(1)_{\rm Y}$. Here the colour
$\SU(3)_{\rm C}$ emerges as the diagonal subgroup of $\SU(3)_{\rm H}$ and the
Georgi-Glashow embedded $\SU(3)\subset\SU(5)$. Likewise, the hypercharge
$\U(1)_{\rm Y}$ is a linear combination of $\U(1)_{\rm H}$ and the usual
hypercharge generator. Adding a pair $T,\ol T$ of Higgs triplet partners in the
${\bf 3}\oplus\ol{\bf 3}$ of $\SU(3)_{\rm H}$, the superpotential terms
\be
W= H\ol T\,\ol Y+\ol HTY
\ee
give GUT-scale masses to the triplet components of $H$ and $\ol H$. Take the
hypercharge generator to be
\be
{\rm Y}={\rm T}^{24}+{\rm Q}\,,
\ee
where ${\rm Q}$ is the $\U(1)_{\rm H}$ charge, and ${\rm
T}^{24}=\sqrt{\frac{3}{5}}\;\diag\,(\frac{1}{3},\frac{1}{3},\frac{1}{3},-\frac{1
}{2},-\frac{1}{2})$. The requirement of leaving hypercharge unbroken then fixes 
the GUT-normalized $\U(1)_{\rm H}$ charge of $Y$ to be ${\rm
Q}[Y]=-\frac{1}{3}$. 

The model also contains a singlet $S$ (which is needed to give a VEV to $Y$ and
$\ol Y$) and an $\SU(3)_{\rm H}$ adjoint $X$ (which is needed to give masses to
the $\SU(3)_{\rm C}$ octet contained in $Y\ol Y$). The particle content is
summarized in Table \ref{tab:pguparticles}. The superpotential is
\be
W=H\ol T\,\ol Y+\ol HTY+S(Y\ol Y+T\ol T-v^2)+Y X\ol Y+T X\ol T+\;(\text{MSSM
Yukawa couplings})\,.
\ee
These are all the renormalizable terms allowed by a certain discrete 
$R$-symmetry, which also serves to forbid dangerous dimension-5 operators.

\begin{table}
\begin{center}
\begin{tabular}{c|c|c|c}
 field & $\SU(5)$ & $\SU(3)_{\rm H}$ & $\U(1)_{\rm H}$ \\ \hline
$3\times{\bf 10}$ & ${\bf 10}$ & ${\bf 1}$ & $0$\\
$3\times\ol{\bf 5}$ & $\ol{\bf 5}$ & ${\bf 1}$ & $0$\\
$H$ & ${\bf 5}$ & ${\bf 1}$ & $0$\\
$\ol H$ & $\ol{\bf 5}$ & ${\bf 1}$ & $0$\\
$T$ & ${\bf 1}$ & ${\bf 3}$ & $1/3$\\
$\ol T$ & ${\bf 1}$ & $\ol{\bf 3}$ & $-1/3$\\
$Y$ & ${\bf 5}$ & $\ol{\bf 3}$ & $-1/3$\\
$\ol Y$ & $\ol{\bf 5}$ & ${\bf 3}$ & $1/3$\\
$S$& ${\bf 1}$ & ${\bf 1}$ & $0$\\
$X$& ${\bf 1}$ & ${\bf 8}$ & $0$\\
\end{tabular}
\end{center}
\caption{Field content of a minimal model with $\SU(5)\times\U(3)$ product group
unification.}\label{tab:pguparticles}.
\end{table}

To extend this to a model of messenger gauge mediation, we add $N_5$ pairs of
messengers $\Phi_I,\,\ol\Phi_I$ in the ${\bf 5}\oplus\ol{\bf 5}$ of $\SU(5)$, 
along with $N_{3{\rm H}}$ pairs of additional fields $\Psi_i,\,\ol\Psi_i$ in 
the ${\bf 3}\oplus\ol{\bf 3}$ of $\SU(3)_{\rm H}$, where $N_5>N_{3{\rm H}}$. 
The $\U(1)_{\rm H}$ charge assignments are arbitrary so far; we take
$Q[\Phi_I]=0$ and $Q[\Psi_i]=\frac{1}{3}$ for simplicity, to match the charges 
of $H$ and $T$. The superpotential operators
\be
W= \lambda_{Ij}\,\Phi_I\ol\Psi_j\,\ol Y+\lambda'_{Ij}\ol\Phi_I\Psi_j Y
\ee
will decouple $N_{3{\rm H}}$ triplets at the scale $M_{\rm GUT}$, leaving
$N_2\equiv N_5$ doublet pairs and $N_3\equiv N_5- N_{3{\rm H}}$ triplet pairs
massless. If $Z=M+F\theta^2$ is a SUSY-breaking spurion, the coupling
\be\label{eq:spurioncoupling}
W= \kappa_{IJ}\,Z\Phi_I\ol\Phi_J
\ee
will eventually give supersymmetric masses $M$ to all remaining messengers. 
We assume that $M\ll M_{\rm GUT}$, such that the GUT-scale massive triplets can be
neglected for SUSY breaking mediation, and $F\ll M^2$ (a possible dynamical 
origin of $M$ and $F$ is sketched later in section \ref{sec:spurion}).
This defines a simple model of messenger gauge mediation, with the somewhat unusual 
property that the light messenger fields do not come in complete GUT multiplets. 
Table\,\ref{tab:messengers} summarizes the messenger field content.

Below the GUT-breaking scale, the messenger couplings become
\begin{eqnarray}
\label{eq:incomplete}
 W = \sum_{i=1}^{N_2}\,\kappa_{2 i}\, Z \Phi_{2i}\ol{\Phi}_{2i} +
  \sum_{a=1}^{N_3}\,\kappa_{3 a}\, Z \Phi_{3a}\ol{\Phi}_{3a}\ ,
\end{eqnarray}
where $\Phi_{2i}$ and $\Phi_{3a}$ denote the remaining $\SU(2)_{\rm L}$ doublet 
and $\SU(3)_{\rm C}$ triplet messengers. 
Here, we have diagonalized the Yukawa interactions (i.e.~the mass matrix
of the remaining messenger fields) at leading order without loss of generality.
It should be noted that the gauge-mediated MSSM soft masses are 
independent of the Yukawa couplings $\k_{2,3}$ at leading order, since 
we are assuming that the messengers couple to a single spurion
and contributions from the GUT-scale messengers are suppressed.

\begin{table}
\begin{center}
\begin{tabular}{c|c|c|c}
 field & $\SU(5)$ & $\SU(3)_{\rm H}$ & $\U(1)_{\rm H}$ \\ \hline
$N_5\times\Phi$ & ${\bf 5}$ & ${\bf 1}$ & $0$\\
$N_5\times\ol\Phi$ & $\ol{\bf 5}$ & ${\bf 1}$ & $0$\\
$N_{3{\rm H}}\times \Psi$ & ${\bf 1}$ & ${\bf 3}$ & $1/3$\\
$N_{3{\rm H}}\times\ol\Psi$ & ${\bf 1}$ & $\ol{\bf 3}$ & $-1/3$\\
\end{tabular}
\end{center}
\caption{Messenger fields and their charges.}\label{tab:messengers}.
\end{table}

We can now investigate the evolution of the gauge couplings, using one-loop
running and step-function decoupling for a rough estimate. We need to consider
three distinct regimes. Below $M$, the field content is that of the MSSM, and
the couplings run as usual. Between $M$ and $M_{\rm GUT}$, the one-loop $\beta$
function coefficients are
\be\label{eq:running1}
b_1=\frac{3}{5}\left(11+N_2+\frac{2}{3} N_3\right)\,,\qquad b_2=1+N_2\,,\qquad
b_3=-3+N_3\,.
\ee
Finally, above $M_{\rm GUT}$ the $\beta$ function coefficients are
\be\begin{split}\label{eq:running2}
b_{1{\rm H}}&=\frac{3}{5}\left(4+\frac{2}{3} N_{3{\rm
H}}\right)=\frac{12}{5}+\frac{2}{5}\left(N_2-N_3\right)\,,\\
b_{3{\rm H}}&=N_{3{\rm H}}=N_2-N_3\,,\\
b_5&=N_5-5=N_2-5\,.
\end{split}
\ee
Furthermore, at $M_{\rm GUT}$ the couplings satisfy the tree-level matching
conditions
\be\label{eq:matching}
\frac{1}{\alpha_1}=\frac{1}{\alpha_{1{\rm
H}}}+\frac{1}{\alpha_5}\,,\qquad\frac{1}{\alpha_2}=\frac{1}{\alpha_5}\,,
\qquad\frac{1}{\alpha_3}=\frac{1}{\alpha_{3{\rm H}}}+\frac{1}{\alpha_5}\,.
\ee
From Eq.~\eqref{eq:matching} it is evident that a unified gauge coupling,
$\alpha_1=\alpha_2=\alpha_3$ at $M_{\rm GUT}$, would correspond to
strongly coupled $\SU(3)_{\rm H}$ and $\U(1)_{\rm H}$ groups at the GUT-breaking
scale. Conversely, if gauge coupling unification is sacrificed by allowing for
nonzero $N_5$ and $N_{3{\rm H}}$, Eqs.~\eqref{eq:running1}, \eqref{eq:running2}
and \eqref{eq:matching} can be used to estimate the scale at which the theory 
becomes strongly coupled in the UV.

\section{A concrete model}

When including a large number of charged fields, it is difficult to construct a
model which remains perturbative all the way to the Planck scale. However, in 
our model a somewhat lower cutoff scale $M_*<M_{\rm Planck}$ is actually 
preferable for a number of reasons. First, a cutoff scale around $M_*=10^{17}$ 
GeV would be of the right order to explain the lack of $m_s-m_\mu$ unification, 
as the Yukawa couplings are corrected by higher-dimensional operators such as 
$W={\bf 10}\,\ol{\bf 5}\,\ol H\,\ol Y\,Y/(M_*)^2$\,.%
\footnote{
At the same order, the term $\ol Y\,Y/(M_*)^2$ can appear
in the gauge kinetic functions, which slightly modifies the matching conditions 
for the gauge coupling constants Eq.~(\ref{eq:matching}).
For $M_*\simeq 10^{17}$\,GeV and ${\cal O}(1)$ coefficients, however, the 
correction to the matching conditions is negligibly small. 
}
And second, a sub-Planckian cutoff allows for a solution of the Polonyi problem 
(which is generally a concern for high-scale gauge mediation, as it is for
gravity mediation; see e.g.~\cite{Ibe:2006am}) using the mechanism of adiabatic
suppression \cite{Nakayama:2011zy}.  

We therefore choose the cutoff scale to be $M_*=10^{17}$ GeV. Moreover, we take
the GUT-breaking scale to be $M_{\rm GUT}=10^{16}$ GeV, and the messenger scale
to be $M=10^{14}$ GeV --- note that a large separation between the messenger
scale and the GUT-breaking scale is preferred, because we will neglect any
contributions to the soft terms from GUT-scale massive triplet messengers.%
\footnote{
Here, we have tacitly rescaled the spurion 
to absorb the typical size of the $\k$ in Eq.\,(\ref{eq:incomplete}), so that 
the messenger scale is given by $M$.
} 
 
In our model $\tan\beta$ is large, and $\mu$ is smaller than the soft
SUSY-breaking terms, because $\mu$ and $B_\mu$ are only generated by subdominant
gravity-mediated effects. In terms of the running parameters at the soft mass 
scale $M_{\rm IR}$, large $\tan\beta$ implies
\be
m_Z^2=-2\left.\left(\,|\mu|^2+\,m_{H_u}^2\right)\right|_{M_{\rm IR}}\,,
\ee 
and thus to obtain a realistic electroweak scale, the contributions to
$m_{H_u}^2$ from the various soft terms have to approximately cancel out in the
renormalization group evolution. This will at most happen for a few select
choices of messenger indices $N_2$ and $N_3$. Numerically solving the two-loop
renormalization group equations between $M_{\rm UV}=10^{14}$ GeV and $M_{\rm
IR}=5\times 10^3$ GeV yields 
\be\begin{split}\label{eq:mHusq}
m_{H_u}^2\Bigr|_{M_{\rm
IR}}=&\Bigl(-0.79\,M_3^2+0.20\,M_2^2-0.01\,M_1\,M_3-0.06\,M_2\,M_3\\
&-0.02\,m_{d_3}^2-0.32\,m_{u_3}^2-0.29\,m_{Q_3}^2+0.04\,m_{H_d}^2+0.70\,m_{H_u}
^2\Bigr)\Bigr|_{M_{\rm UV}}\,.
\end{split}
\ee
Here we have omitted terms with coefficients $<0.01$ (although they are
internally kept in the following calculations). We have also neglected any terms
involving the $A$-parameters at the messenger scale, since these are expected to
be small in gauge mediation. The standard one-loop messenger gauge mediation
expressions for the gaugino masses at $M_{\rm UV}=M$ are
\be
\begin{split}\label{eq:gmsb1}
 M_1=&\;\frac{g_1^2}{16\pi^2}\frac{F}{M}\left(\frac{3}{5}N_2+\frac{2}{5}
N_3\right)\,,\\
M_2=&\;\frac{g_2^2}{16\pi^2}\frac{F}{M}\,N_2\,,\\
M_3=&\;\frac{g_3^2}{16\pi^2}\frac{F}{M}\,N_3\,,
\end{split}
\ee
while the scalar soft masses are
\be
\begin{split}\label{eq:gmsb2}
m_{Q_3}^2=&\;2\left(\frac{F}{M}\right)^2\left[\left(\frac{g_3^2}{16\pi^2}
\right)^2\cdot\frac{4}{3}N_3+\left(\frac{g_2^2}{16\pi^2}\right)^2\cdot\frac{3}{4
}N_2+\left(\frac{g_1^2}{16\pi^2}\right)^2\cdot\frac{1}{60}\left(\frac{2}{5}
N_3+\frac{3}{5}N_2\right)\right]\,,\\
m_{u_3}^2=&\;2\left(\frac{F}{M}\right)^2\left[\left(\frac{g_3^2}{16\pi^2}
\right)^2\cdot\frac{4}{3}N_3+\left(\frac{g_1^2}{16\pi^2}\right)^2\cdot\frac{4}{
15}\left(\frac{2}{5}N_3+\frac{3}{5}N_2\right)\right]\,,\\
m_{d_3}^2=&\;2\left(\frac{F}{M}\right)^2\left[\left(\frac{g_3^2}{16\pi^2}
\right)^2\cdot\frac{4}{3}N_3+\left(\frac{g_1^2}{16\pi^2}\right)^2\cdot\frac{1}{
15}\left(\frac{2}{5}N_3+\frac{3}{5}N_2\right)\right]\,,\\
m_{H_u}^2=&\;m_{H_d}^2=2\left(\frac{F}{M}\right)^2\left[\left(\frac{g_2^2}{
16\pi^2}\right)^2\cdot\frac{3}{4}N_2+\left(\frac{g_1^2}{16\pi^2}
\right)^2\cdot\frac{3}{20}\left(\frac{2}{5}N_3+\frac{3}{5}N_2\right)\right]\,.
\end{split}
\ee
Crucially, as mentioned before and as usual in minimal gauge mediation, the soft 
terms only depend on the ratio $F/M$ and on the messenger field content. In 
particular there is no dependence on the unknown Yukawa couplings $\kappa$ in
Eqs.~\eqref{eq:spurioncoupling} or \eqref{eq:incomplete}.%
\footnote{
Strictly speaking, the above soft masses are valid only for degenerate couplings 
$\k_{2,i} = \k_{3,a}$, whereupon all the messengers decouple at the same 
messenger scale. However, our results are not significantly altered even if the
$\k$ are not all equal, as long as they are of a similar order of magnitude.
}

Substituting Eqs.~\eqref{eq:gmsb1} and
\eqref{eq:gmsb2} into Eq.~\eqref{eq:mHusq} we obtain
\be
m_{H_u}^2\Bigr|_{M_{\rm
IR}}=\left(\frac{1}{16\pi^2}\frac{F}{M}\right)^2\left(-0.215\,N_3^2-0.449\,
N_3+0.044\,N_2^2+0.150\,N_2-0.015\,N_3\,N_2\right)\,.
\ee
Suitable combinations for $(N_2,N_3)$ to ensure small negative $m_{H_u}^2$ at
$M_{\rm IR}$ are given by $(N_2,N_3)=(17,7)$ or $(N_2, N_3)=(22,9)$. (Note that
the ratio $N_3\;:\:N_2$
is always around $2\;:\;5$ for a ``gaugino focus point'', for $N_2$ and $N_3$
sufficiently large. Here, both $7/17$ and $9/22$ are close to $0.41$.)

\begin{figure}
\centering
 \includegraphics[width=.6\textwidth]{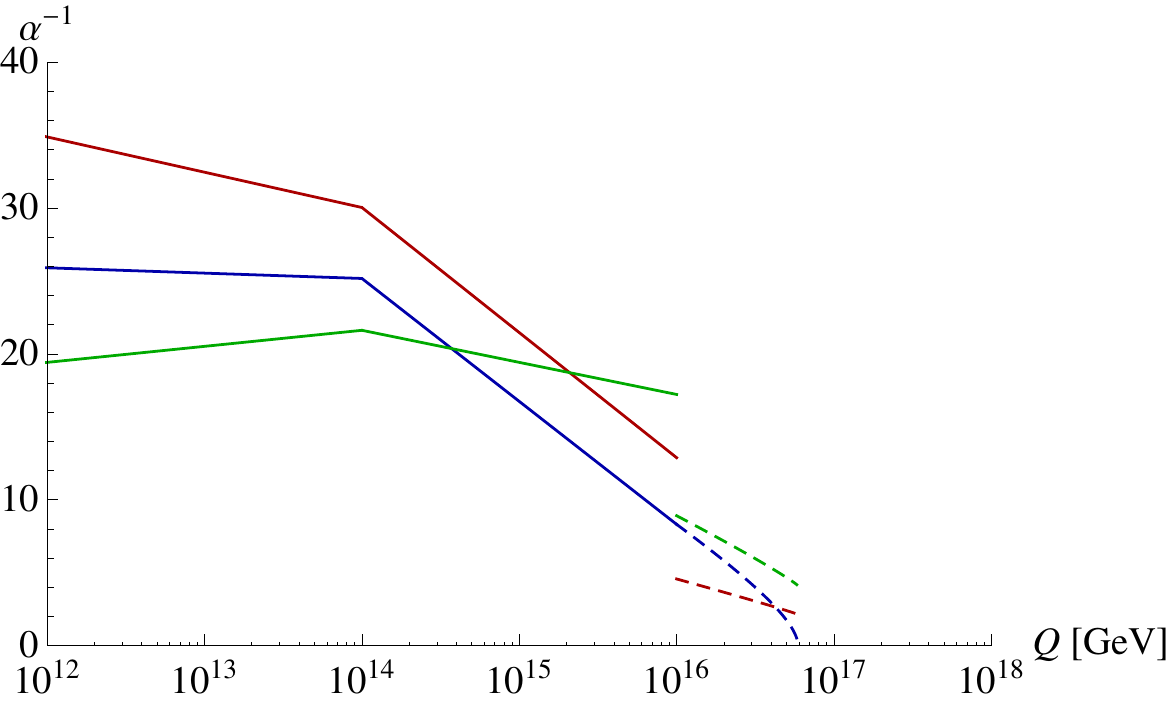}
\caption{Evolution of the gauge couplings for
$(N_2,N_3)=(22,9)$. Solid red curve: $\alpha_1^{-1}$, solid blue curve:
$\alpha_2^{-1}$,
solid green curve: $\alpha_3^{-1}$. Dashed red curve: $\alpha_{1{\rm H}}^{-1}$,
dashed blue curve: $\alpha_5^{-1}$, dashed green curve: $\alpha_{3{\rm
H}}^{-1}$. Two-loop running is used above the matching scale $M_{\rm
GUT}=10^{16}$ GeV, but the effect of Yukawa couplings is neglected. 
}\label{fig:running5}
\end{figure}

\begin{figure}
\centering
 \includegraphics[width=.6\textwidth]{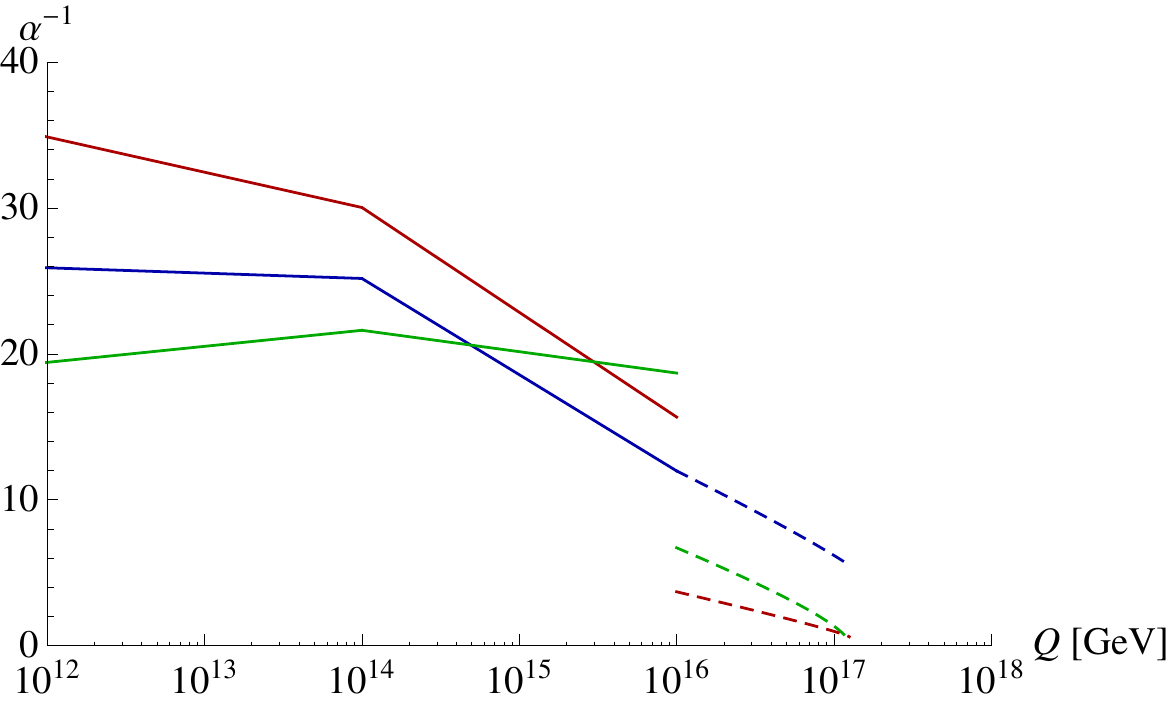}
\caption{Evolution of the gauge couplings for
$(N_2,N_3)=(17,7)$. The colour code is the same as in Fig.~\ref{fig:running5}. 
}\label{fig:running6}
\end{figure}

The one-loop running of the gauge couplings is shown in Fig.~\ref{fig:running5}
for 
$(N_2,N_3)=(22,9)$ and in Fig.~\ref{fig:running6} for  $(N_2,N_3)=(17,7)$. Above
the GUT-breaking scale the couplings are seen to be quite large, and the
reliability of the one-loop approximation should be questioned. We have
therefore 
used two-loop running in this regime, in the (conservative) limit where any 
Yukawa couplings can be neglected. 

In the $(N_2,N_3)=(22,9)$ model, the two-loop correction is quite significant, 
and leads to the theory becoming perturbatively unreliable at scales below the 
previously assumed cutoff $M_*$. We therefore discard this possibility: the
matter content of the model is too large to guarantee perturbative control an 
order of magnitude above the GUT-breaking scale. Note, however, that sizeable 
Yukawa couplings might still change this behaviour.

On the other hand, in the $(N_2,N_3)=(17,7)$  model a Landau pole is reached
only around $M_*=10^{17}$ GeV. We therefore choose this model to compute a 
benchmark parameter point using \verb!SOFTSUSY! \cite{Allanach:2001kg}. Set 
$F=(4\times 10^9\,\text{GeV})^2$, 
which implies that $\mu$ and $\sqrt{B_\mu}$ should be ${\cal O}(F/M_*)\sim 200$
GeV. For practical purposes, they are fixed by 
requiring realistic electroweak symmetry breaking. In our benchmark point 
they come out slightly high: $\mu(M)=232$ GeV and $\sqrt{B_\mu}(M)=614$ GeV. 
For scalar and gaugino soft masses, we use only the gauge-mediated 
contributions from Eqs.~\eqref{eq:gmsb1} and \eqref{eq:gmsb2}, disregarding 
possible ``gravity-mediated'' corrections ${\cal O}(F/M_*)$.

The low-energy sparticle spectrum is shown in Table \ref{tab:spectrum}. Note 
that the lightest Higgs mass is on the low end of what is compatible with the 
LHC observation, when taking a theory uncertainty of $3$ GeV into account. The 
reason is that, even with multi-TeV stops, a large Higgs mass is difficult
to obtain if the $A$-terms are only generated radiatively. The superpartners are
seen to be too heavy to be produced at the LHC as expected, with the exception 
of some relatively light higgsino-like charginos and neutralinos $\chi^\pm_1$,
$\chi^0_1$, 
and $\chi^0_2$. This is similar to previous models of high-scale gauge mediation
in which the $\mu$ term is induced by subdominant gravity mediation effects, see 
e.g.~\cite{Brummer:2011yd}. Light higgsinos will be difficult to see at the LHC, 
but could be discovered at a future linear collider. 

\begin{table}
\begin{center}
\begin{tabular}[t]{c|c} particle &  mass [GeV]\\ \hline
$h_0$ & $123$\\ 
$\chi^0_1$ &  $207$ \\ 
$\chi^\pm_1$ & $208$  \\ 
$\chi^0_2$ & $209$ \\ 
$\chi^0_3$ &  $2900$ \\ 
$\chi^0_4$ & $6900$ \\ 
$\chi^\pm_2$ & $6900$  \\ 
$H_0$ & $3000$ \\ 
\end{tabular}
\hspace{2cm}
\begin{tabular}[t]{c|c} particle &  mass [GeV]\\ \hline
$A_0$ & $3000$ \\ 
$H^\pm$ & $3000$ \\ 
$\tilde g$ & $7000$ \\ 
$\tilde\tau_1$ & $2000$ \\
other sleptons & $3200 - 6000$\\ 
$\tilde t_1$ & $5000$ \\
$\tilde t_2$ & $7500$ \\ 
other squarks & $6400 - 8600$ \\
\end{tabular}
\caption{Mass spectrum for $(N_2,N_3)=(17,7)$, with $F=(4\times 10^9\text{
GeV})^2$, $M=10^{14}$ GeV, $\mu=230$ GeV, and $\sqrt{B_\mu}=614$ GeV. For these
parameters, $\tan\beta=49$.}\label{tab:spectrum}
\end{center}
\end{table}

To estimate the fine-tuning, we take the usual definition
\be
\Delta=\max_{\text{parameters }a}\frac{\partial\log m_Z}{\partial\log
a}\,,\qquad\qquad\text{fine-tuning}=\frac{1}{\Delta}\,.
\ee
There are only two independent dimensionful parameters $a$ which enter $m_Z$ (at
large $\tan\beta$). Namely,
\be
m_Z^2=-2\left.\left(|\mu|^2+m_{H_u}^2\right)\right|_{M_{\rm IR}}=0.42\,m_{\rm
GMSB}^2 - 1.62\,|\mu|^2\Bigr|_{M_{\rm UV}}
\ee
where 
\be
m_{\rm GMSB}\equiv\frac{1}{16\pi^2}\frac{F}{M}\approx 1\;\TEV\,.
\ee
While this relation is not precise enough to predict $m_Z$ accurately, it does
serve to illustrate that the residual fine-tuning is of the order $(m_Z/m_{\rm
GMSB})^2$ (because the coefficient of $m_{\rm GMSB}$ is ${\cal O}(1)$). For the
present benchmark point, this implies that $1/\Delta\sim {\cal O}(1\,\%)$,
which is a considerable improvement over generic models with similarly heavy 
superpartners. It should be pointed out, however, that the sensitivity of the 
focus point cancellation to the dimensionless Standard Model couplings (which 
are usually not included into the fine-tuning definition) is likely very high. 

One might object that the particular choice of messenger content in our model
constitutes a fine-tuning by itself which we should take into account. Our
take is that the field content of a given model should be regarded as part of 
the model definition, and therefore not included in the fine-tuning computation.
(Note also that the canonical quantitative measure of fine-tuning,  the
sensitivity of the weak scale with respect to parameter variations, is
ill-defined for ``discrete parameters'' such as the number of particle
flavours.) While it is meaningful to compare different models, defined by
different symmetries and field content, with regards to their fine-tuning in
their respective continuous parameters, there is no natural measure which a
priori prefers any particular model over another. At best, we can aim for
simplicity and minimality. By these standards our peculiar messenger field
content may admittedly be considered a drawback of our model --- as often in
model-building, reduced fine-tuning comes at the price of a more elaborate
construction. However, the fairly large messenger numbers in our
model are also a consequence of demanding that the gauge couplings should
be perturbative above the GUT-breaking scale. This condition thus serves as
an independent motivation for considering a model in which $N_2$ and $N_3$
are of the order 10--20. 

Finally, if the $Z$ superfield of the previous section is identified with the goldstino
multiplet, then the LSP is a $4$ GeV gravitino. 
If furthermore $R$-parity is conserved, within standard cosmology, the decays of the 
$\chi^0_1$ NLSP will spoil the successful prediction of light element abundances during Big Bang 
Nucleosynthesis. 
In the current model, however, the relic abundance of the NSLP is rather low,
$\Omega_\chi h^2 \lesssim 10^{-2}$, due to the large annihilation cross section
of the Higgsino. Therefore the BBN problem can solved by slightly lowering the SUSY 
breaking, such as to lower the gravitino mass to $\lesssim 1$ GeV which reduces 
the higgsino lifetime to ${\cal O}(10^2)$ seconds
\cite{Kawasaki:2004qu,Jedamzik:2006xz}.%
\footnote{Note however that the gravitino cannot be made arbitrarily light.
To obtain an MSSM spectrum similar to the one of Table~\ref{tab:spectrum},
the messenger scale also would need to be lowered accordingly, keeping $F/M$ 
constant. But a too low messenger scale would result in a too large $g_5$ coupling 
at $M_{\rm GUT}$, which would then again blow up very quickly.}
On the other hand, the actual goldstino direction of the hidden sector may have other
components besides $Z$ (as in the model of the following section), in which case the 
gravitino could also be heavier than $\chi^0_1$.%
\footnote{
In the Higgsino LSP case, the thermal relic abundance is too small to account for the 
observed dark matter density, so one would need non-thermal sources for the Higgsino
or another dark matter candidate.
}

\section{The origin of the SUSY breaking spurion}\label{sec:spurion}
So far, we have simply assumed that the messengers couple to a SUSY-breaking 
spurion field $Z$ whose vacuum expectation value is
\begin{eqnarray}
\vev{Z} = M + F \theta^2\ .
\end{eqnarray}
In this section, we sketch a possible dynamical origin of $F$ and $M$.
We consider the model of cascade SUSY breaking which was developed in 
Refs.~\cite{Ibe:2010jb,Evans:2011pz}. This model contains a primary SUSY-breaking
field $Z_0$ and a secondary SUSY-breaking field $Z_1$, the latter of which will
be identified with $Z$.

In the cascade SUSY breaking model, the K\"ahler potential and superpotential are
\begin{eqnarray}
\label{eq:cascade}
K &=& Z_0^\dagger Z_0 +Z_1^\dag Z_1
- \frac{c_0^2}{4\Lambda^2}\, (Z_0^\dagger Z_0)^2
+ \frac{c_1^2}{\Lambda^2} Z_0^\dagger Z_0  Z_1^\dagger Z_1  + \ldots\ , \\
W &=& \Lambda^2 Z_0 + \frac{h}{3} Z_1^3 + \kappa\,Z_1 \Phi \ol{\Phi}\,,
\end{eqnarray}
where $\Lambda$ denotes a dimensionful parameter while $c_{0,1}$, $h$ and 
$\kappa$ are dimensionless coefficients.
In the following, we take $h$ and $\kappa$ to be real and positive by
rotating the phases of the fields appropriately.
We are assuming that the higher dimensional  K\"ahler potential terms are 
generated radiatively by integrating out certain fields in a generalized 
O'Raifeartaigh model \cite{Ibe:2010jb,Evans:2011pz}. It can be shown that
$c_0^2>0$ when the quartic $Z_0$ term is perturbatively generated by 
integrating out fields with $R$-charges $0$ or $2$\,\cite{Shih:2007av}.
Similarly, for $c_1^2 > 0$ one needs $Z_1$ to couple to fields with 
$R$-charges other than $0$ or $2$\,\cite{Ibe:2010jb,Evans:2011pz}.

In this model, there is an $R$-symmetric but SUSY-breaking vacuum at
\begin{eqnarray}
 Z_0 &=& 0\,, \\
 F_{Z_0} &=& \L^2\,,
\end{eqnarray}
around which the primary SUSY-breaking field obtains a mass
\begin{eqnarray}
m_{Z_0}^2 = c_0^2  \Lambda^2 \ .
\end{eqnarray}

Once $Z_0$ breaks SUSY, 
the secondary SUSY-breaking field $Z_1$ obtains a soft SUSY-breaking mass term.
The scalar potential for $Z_1$ is given by 
\begin{equation}
V(Z_1) \simeq  - m_{Z_1}^2 |Z_1|^2 + | h Z_1^2|^2,\qquad 
m_{Z_1}^2 = c_1^2 \frac{|F_{Z_0}|^2}{\L^2}
= c_1^2 \L^2
\ .
\end{equation}
Therefore, for $m_{Z_1}^2 > 0$, the secondary 
SUSY-breaking field obtains a non-vanishing expectation value,
\begin{eqnarray}
\label{eq:vevSapp}
 \vev {Z_1} \simeq \frac{m_{Z_1}}{\sqrt 2 h}
 \simeq \frac{c_1}{\sqrt 2 h} \L
 \ ,
\end{eqnarray}
which breaks SUSY by
\begin{eqnarray}
\label{eq:vevFapp}
 F_{Z_1} =  h\vev{Z_1^*}^{2} \simeq  \frac{m_{Z_1}^2}{2 h}
 =  \frac{c_1^2}{2 h}\L^2 \ .
\end{eqnarray}
In this way, secondary SUSY breaking
is initiated by spontaneous $R$-symmetry breaking which is, in turn, triggered by
fundamental SUSY breaking.%

Through the coupling between the messengers and $Z_1$ in Eq.\,(\ref{eq:cascade}),
the secondary SUSY breaking field $Z_1$ plays the role of the spurion 
in the previous sections, i.e.
\begin{eqnarray}
M = \kappa\,\vev{Z_1} \simeq \frac{\kappa c_1}{\sqrt{2}h} \L\ , \quad
F= \kappa\,F_{Z_1} \simeq  \frac{\kappa c_1^2}{2 h}\L^2 \ .
\end{eqnarray}
In the following, we assume that $c_1^2 \simeq 2 h$ and $\kappa \simeq 1$, 
in order to obtain a gravitino mass which is
as low as possible for a given $F$, i.e. $F \simeq  F_{Z_1} \simeq F_{Z_0}$,
while keeping the messenger mass as high as possible.%
\footnote{
As mentioned above, another possibility would be to have a much heavier
gravitino mass and a Higgsino LSP.
In this case, parameter choices such as $c_1\ll h$ and $\k\ll 1$ are also allowed.
}
Under these assumptions, the parameters used in the previous section, 
\begin{eqnarray}
 M \simeq 10^{14}\,{\rm GeV}\ , \quad  F\simeq (4\times 10^9\,\text{GeV})^2\ ,
\end{eqnarray}
are obtained by choosing
\begin{eqnarray}
 \Lambda \simeq 4\times 10^9\,\text{GeV}\ , \quad c_1 \simeq 5\times 10^{-5 }\ .
\end{eqnarray}

Finally, let us comment on the origins of the $\mu$ and $B_\mu$ term.
As mentioned earlier, one may consider ``gravity-mediated" contributions,
\begin{eqnarray}
 K = c_H\frac{Z^\dagger}{M_*} H_u H_d +c_B \frac{Z^\dagger Z}{M_*^2} H_u H_d\hc \ ,
\end{eqnarray}
which leads to
\begin{eqnarray}
\mu &=& \sqrt{3}c_H \frac{M_{\rm Planck}} {M_*} m_{3/2} \ , \\
\sqrt{B_\mu} &=& \sqrt{3}c_B^{1/2} \frac{M_{\rm Planck}} {M_*} m_{3/2} \ .
\end{eqnarray}
Thus, with the coefficients $c_H$ and $c_B$ of order unity and 
$M_* \simeq 10^{17}$\,GeV, one obtains values for $\mu$ and $B_\mu$ of the 
order of the weak scale as desired.

It should be noted that one of the phases of $c_{H,B}$ cannot be eliminated 
by field redefinitions, which leads to CP-violating processes and gives rise to, 
for instance, an electron electric dipole moment (EDM).
In fact, if the relative phase between $\mu$ and $B_\mu$ is of order unity,
the predicted electron EDM slightly exceeds the current limit
for the mass spectrum in Table\,\ref{tab:spectrum}. Avoiding the bound
requires some amount of tuning between $c_H$ and $c_B$.

Similarly, completely generic gravity-mediated contributions to soft masses 
and $A$-terms would lead to unacceptably large flavour changing neutral 
currents. While the soft terms are dominated by the flavour-universal
gauge-mediated contributions, subdominant flavour-violating corrections are 
still potentially dangerous, see e.g.~\cite{Hiller:2008sv,Hiller:2010dv}. 
Keeping them under control also requires some tuning or an underlying symmetry. 
However, a detailed analysis of the flavour and CP problems in our model is 
beyond the scope of this work.

\section{Conclusions}

We have shown that focus point gauge mediation can naturally be embedded into
a model of $\SU(5)\times\U(3)_{\rm H}$ product group unification. For the cutoff of the
product group unification model to be substantially higher than the GUT-breaking
scale, we have added to the MSSM a number of colour triplet and weak doublet 
pairs at intermediate energies. In terms of full $\SU(5)\times\U(3)_{\rm H}$ 
representations, this corresponds to adding several ${\bf 5}\oplus\bar{\bf 5}$ 
and ${\bf 3}\oplus\bar{\bf 3}$ pairs. Importantly, if the cutoff of the 
model is to be significantly beyond the GUT-breaking scale, this implies that
the number of extra states should be fairly large. For instance, with $N_2=17$ 
extra doublet pairs and $N_3=7$ extra triplet pairs, the cutoff can be postponed
until $M_*\approx 10^{17}$ GeV. If the number of extra states is further
increased (as in the $(N_2,N_3)=(9,22)$ model which we also discussed), a Landau 
pole in the $\SU(5)$ gauge coupling appears just above the GUT-breaking scale. Too 
few extra states would also imply a lower cutoff scale, as can be seen from the 
extreme case $(N_2,N_3)=(0,0)$ where the $\U(1)_{\rm H}$ gauge coupling would have to be 
non-perturbative already at $M_{\rm GUT}$. 

The observation that both $\U(3)_{\rm H}$ couplings become strong at 
$M_*\approx 10^{17}$ GeV may indicate that the $\U(3)_{\rm H}$ gauge bosons are composite
states at the cutoff scale $M_*$ \cite{Landau:1954, Landau:1955ip, Landau:1955}. 
To realize a compositeness scale between the GUT scale and the Planck scale,
introducing additional charged states, as we have done, in fact becomes a necessity. 
The $\SU(5)$ gauge bosons, on the other hand, may be elementary
(or they might be composite as well, if further ${\bf 5}\oplus\bar{\bf 5}$ 
multiplets with GUT-scale masses are introduced).
 
The extra weak doublets and colour triplets act as gauge mediation messengers, 
leading to the non-unified soft term mass ratios which are required for a focus 
point-like cancellation between the radiative corrections to the Higgs potential, 
and consequently a little hierarchy between the soft mass scale and the electroweak 
scale. The superpartners, except for some higgsino-like neutralinos and charginos, 
are predicted to be very heavy and out of LHC reach. The resulting Higgs mass is 
compatible with 125 GeV, while the fine-tuning is still comparatively modest.

\section*{Acknowledgments}
This work is supported by Grant-in-Aid for Scientific research from the
Ministry of Education, Science, Sports, and Culture (MEXT), Japan, No.\ 22244021 (T.T.Y.),
No.\ 24740151 (M.I.), and also by the World Premier International Research Center Initiative (WPI Initiative), MEXT, Japan.


\begin{thebibliography}{99}
\bibitem{Abe:2007kf}
  H.~Abe, T.~Kobayashi and Y.~Omura,
  Phys.\ Rev.\ D {\bf 76} (2007) 015002
  [hep-ph/0703044 [HEP-PH]].

\bibitem{Horton:2009ed}
  D.~Horton and G.~G.~Ross,
  Nucl.\ Phys.\ B {\bf 830} (2010) 221
  [arXiv:0908.0857 [hep-ph]].

\bibitem{Brummer:2012zc}
  F.~Br\"ummer and W.~Buchm\"uller,
  JHEP {\bf 1205} (2012) 006
  [arXiv:1201.4338 [hep-ph]].

\bibitem{Brummer:2011yd}
  F.~Br\"ummer and W.~Buchm\"uller,
  JHEP {\bf 1107} (2011) 010
  [arXiv:1105.0802 [hep-ph]].

\bibitem{Younkin:2012ui}
  J.~E.~Younkin and S.~P.~Martin,
  Phys.\ Rev.\ D {\bf 85} (2012) 055028
  [arXiv:1201.2989 [hep-ph]].

\bibitem{Antusch:2012gv}
  S.~Antusch, L.~Calibbi, V.~Maurer, M.~Monaco and M.~Spinrath,
  arXiv:1207.7236 [hep-ph].
\bibitem{Yanagida:2013ah}
  T.~T.~Yanagida and N.~Yokozaki,
  arXiv:1301.1137 [hep-ph].



\bibitem{Chan:1997bi}
  K.~L.~Chan, U.~Chattopadhyay and P.~Nath,
  Phys.\ Rev.\ D {\bf 58} (1998) 096004
  [hep-ph/9710473].

\bibitem{Feng:1999mn}
  J.~L.~Feng, K.~T.~Matchev and T.~Moroi,
  Phys.\ Rev.\ Lett.\  {\bf 84} (2000) 2322
  [hep-ph/9908309].

\bibitem{Feng:1999zg}
  J.~L.~Feng, K.~T.~Matchev and T.~Moroi,
  Phys.\ Rev.\ D {\bf 61} (2000) 075005
  [hep-ph/9909334].

\bibitem{Yanagida:1994vq}
  T.~Yanagida,
  Phys.\ Lett.\ B {\bf 344} (1995) 211
  [hep-ph/9409329].

\bibitem{Hisano:1995hc}
  J.~Hisano and T.~Yanagida,
  Mod.\ Phys.\ Lett.\ A {\bf 10} (1995) 3097
  [hep-ph/9510277].

\bibitem{Izawa:1997he}
  K.~I.~Izawa and T.~Yanagida,
  Prog.\ Theor.\ Phys.\  {\bf 97} (1997) 913
  [hep-ph/9703350].

\bibitem{Ibe:2006am}
  M.~Ibe, Y.~Shinbara and T.~T.~Yanagida,
  Phys.\ Lett.\ B {\bf 639} (2006) 534
  [hep-ph/0605252].

\bibitem{Nakayama:2011zy}
  K.~Nakayama, F.~Takahashi and T.~T.~Yanagida,
  Phys.\ Rev.\ D {\bf 86} (2012) 043507
  [arXiv:1112.0418 [hep-ph]].

\bibitem{Allanach:2001kg}
  B.~C.~Allanach,
  Comput.\ Phys.\ Commun.\  {\bf 143} (2002) 305
  [hep-ph/0104145].

\bibitem{Kawasaki:2004qu}
  M.~Kawasaki, K.~Kohri and T.~Moroi,
  Phys.\ Rev.\ D {\bf 71} (2005) 083502
  [astro-ph/0408426].

\bibitem{Jedamzik:2006xz}
  K.~Jedamzik,
  Phys.\ Rev.\ D {\bf 74} (2006) 103509
  [hep-ph/0604251].

\bibitem{Ibe:2010jb}
  M.~Ibe, Y.~Shirman and T.~T.~Yanagida,
  JHEP {\bf 1012} (2010) 027
  [arXiv:1009.2818 [hep-ph]].
 
\bibitem{Evans:2011pz}
  J.~L.~Evans, M.~Ibe, M.~Sudano and T.~T.~Yanagida,
  JHEP {\bf 1203} (2012) 004
  [arXiv:1103.4549 [hep-ph]].

\bibitem{Shih:2007av}
  D.~Shih,
  JHEP {\bf 0802} (2008) 091
  [hep-th/0703196].

\bibitem{Ibe:2007km}
  M.~Ibe and R.~Kitano,
  JHEP {\bf 0708} (2007) 016
  [arXiv:0705.3686 [hep-ph]].

\bibitem{Hiller:2008sv}
  G.~Hiller, Y.~Hochberg and Y.~Nir,
  JHEP {\bf 0903} (2009) 115
  [arXiv:0812.0511 [hep-ph]].

\bibitem{Hiller:2010dv}
  G.~Hiller, Y.~Hochberg and Y.~Nir,
  JHEP {\bf 1003} (2010) 079
  [arXiv:1001.1513 [hep-ph]].

\bibitem{Landau:1954}
L.~D.~Landau, A.~A.~Abrikosov and I.~M.~Khalatnikov, Dokl. Akad. Nauk USSR {\bf 95} (1954) 773, 1177;
{\it ibid.}~{\bf 96} (1954) 261.

\bibitem{Landau:1955ip}
  L.~D.~Landau and I.~Y.~.Pomeranchuk,
  Dokl.\ Akad.\ Nauk Ser.\ Fiz.\  {\bf 102} (1955) 489.

\bibitem{Landau:1955}
L.~D.~Landau, ``Niels Bohr and the development of physics,'' ed.~W.~Pauli (Pergamon Press, 1955).



 
\end{thebibliography}
\end{document}